\documentclass[longauth]{aa}  

\usepackage{graphicx}
\usepackage{rotating}
\usepackage{txfonts}
\usepackage{subfigure}
\usepackage{natbib}
\bibpunct{(}{)}{;}{a}{}{,} 

\begin{document}

   \title{The impact of the Star Formation Histories on the  SFR-M$_{*}$ relation at z $\ge$2}
\author{L.~P. Cassar\`{a}\inst{1,19} \and D. Maccagni\inst{1} \and B. Garilli\inst{1}
\and M. Scodeggio\inst{1}
\and R. Thomas\inst{3}
\and O. Le F\`evre\inst{3}
\and G. Zamorani \inst{2}
\and D. Schaerer\inst{10,8}
\and B.C. Lemaux \inst{3}
\and P. Cassata\inst{18}
\and V. Le Brun\inst{3}
\and L. Pentericci\inst{4}
\and L.A.M. Tasca\inst{3}
\and E. Vanzella\inst{2}
\and E. Zucca\inst{2}
\and R. Amor\'in\inst{4}
\and S. Bardelli\inst{2}
\and M. Castellano\inst{4}
\and A. Cimatti\inst{5}
\and O. Cucciati\inst{5,2}
\and A. Durkalec\inst{3}
\and A. Fontana\inst{4}
\and M. Giavalisco\inst{13}
\and A. Grazian\inst{4}
\and N. P. Hathi\inst{3}
\and O. Ilbert\inst{3}
\and S. Paltani\inst{9}
\and B. Ribeiro\inst{3}
\and V. Sommariva\inst{5,4}
\and M. Talia\inst{5}
\and L. Tresse\inst{3}
\and D. Vergani\inst{6,2}
\and P. Capak\inst{12}
\and S. Charlot\inst{7}
\and T. Contini\inst{8}
\and S. de la Torre\inst{3}
\and J. Dunlop\inst{16}
\and S. Fotopoulou\inst{9}
\and L. Guaita\inst{4}
\and A. Koekemoer\inst{17}
\and C. L\'opez-Sanjuan\inst{11}
\and Y. Mellier\inst{7}
\and J. Pforr\inst{3}
\and M. Salvato\inst{14}
\and N. Scoville\inst{12}
\and Y. Taniguchi\inst{15}
\and P.W. Wang\inst{3}
}
\institute{INAF--IASF Milano, via Bassini 15, I-20133, Milano, Italy
\and
INAF--Osservatorio Astronomico di Bologna, via Ranzani,1, I-40127, Bologna, Italy
\and
Aix Marseille Universit\'e, CNRS, LAM (Laboratoire d'Astrophysique de
Marseille) UMR 7326, 13388, Marseille, France
\and
INAF--Osservatorio Astronomico di Roma, via di Frascati 33, I-00040, Monte Porzio
Catone, Italy
\and
University of Bologna, Department of Physics and Astronomy (DIFA), V.le Berti
Pichat, 6/2 - 40127, Bologna, Italy
\and
INAF--IASF Bologna, via Gobetti 101, I--40129, Bologna, Italy
\and
Institut d'Astrophysique de Paris, UMR7095 CNRS,
Universit\'e Pierre et Marie Curie, 98 bis Boulevard Arago, 75014
Paris, France
\and
Institut de Recherche en Astrophysique et Plan\'etologie - IRAP, CNRS, Université de
Toulouse, UPS-OMP, 14, avenue E. Belin, F31400
Toulouse, France
\and
Department of Astronomy, University of Geneva,
ch. d'Ecogia 16, CH-1290 Versoix, Switzerland
\and
Geneva Observatory, University of Geneva, ch. des Maillettes 51, CH-1290 Versoix,
Switzerland
\and
Centro de Estudios de F\'isica del Cosmos de Arag\'on, Teruel, Spain
\and
Department of Astronomy, California Institute of Technology, 1200 E. California
Blvd., MC 249--17, Pasadena, CA 91125, USA
\and
Astronomy Department, University of Massachusetts, Amherst, MA 01003, USA
\and
Max-Planck-Institut f\"ur Extraterrestrische Physik, Postfach 1312, D-85741,
Garching bei M\"unchen, Germany
\and
Research Center for Space and Cosmic Evolution, Ehime University, Bunkyo-cho 2-5,
Matsuyama 790-8577, Japan
\and
SUPA, Institute for Astronomy, University of Edinburgh, Royal Observatory,
Edinburgh, EH9 3HJ, United Kingdom
\and
Space Telescope Science Institute, 3700 San Martin Drive, Baltimore, MD 21218, USA 
\and
Instituto de Fisica y Astronomia, Facultad de Ciencias, Universidad de Valparaiso,
Av. Gran Bretana 1111, Casilla 5030, Valparaiso, Chile
\and
Institute for Astronomy, Astrophysics, Space Applications and Remote
Sensing, National Observatory of Athens, Penteli, 15236, Athens, Greece
}

\date{Accepted for publication in A\&A/ In its original form 2015 May 9} 
				
	\authorrunning{author(s)}
	\titlerunning{title}

 \abstract {In this paper we investigate the impact of different star formation histories (SFHs) on the relation between 
stellar mass (M$_{*}$) and star formation rate
(SFR) using a sample of galaxies with reliable spectroscopic redshift $z_{\rm spec}>2$ drawn from the VIMOS Ultra-Deep Survey (VUDS).
We produce an extensive database of dusty model galaxies, calculated
starting from the new library of single stellar population (SSPs) models presented in \citet{Cassara2013} and weighted
by a set of 28 different star formation histories based on the Schmidt function, and characterized by different
ratios of the gas infall time scale $\tau_{\rm infall}$ to the star formation efficiency $\nu$.
The treatment of dust extinction and re-emission has been carried out by means of the radiative transfer calculation.
The spectral energy distribution (SED) fitting technique is performed by using GOSSIP+, a tool able to combine both photometric and spectroscopic 
information to extract the best value of the physical quantities of interest, and to consider the Intergalactic Medium (IGM) attenuation as a 
free parameter.
We find that the main contribution to the scatter observed in the SFR-M$_{*}$ plane is the possibility of choosing between different families of SFHs in the SED fitting procedure, while the redshift range plays a minor role.
The majority of the galaxies, at all cosmic times, are best-fit by models with SFHs characterized by a high $\tau_{\rm infall}/\nu$ ratio. 
We discuss the reliability of the presence of a small percentage of dusty and highly star forming galaxies, in the light of their detection
in the FIR.
}

												
  \maketitle
%

\section{Introduction}

A fundamental observable in astrophysical cosmology is the cosmic history of
star formation. In a recent review, \citet{Madau2014} note that not only it
is still hard to accurately determine the redshift of maximum star formation rate density (SFRD), but that 
beyond redshift $\sim$ 2 there is considerable uncertainty in the amount of stellar 
light obscured by dust. To this, we can add the uncertainties in deriving 
stellar masses and star formation rates (SFRs) for large complete samples of galaxies with well known
selection functions.
Specifically, reliable indicators of the SFR, like the 
H$\alpha$ flux, are redshifted into the $K$ band or beyond at z>2, where spectrographs
do not yet allow to collect large samples. Other indicators, like the far-infrared (FIR) 
luminosity, are limited to the brightest population because of the sensitivity of Herschel 
\citep{Lemaux2013,Rodighiero2014}.
In the last years there have been some  developments in terms of estimating SFRs on
composite or individual galaxies at higher redshift through Herschel
stacking \citep{Alvarez2016}, through color
evolution/estimating EWs of recombination lines with broadband
photometry \citep{Faisst2016}, VLA/ALMA observations of galaxies
that are star-forming at relatively normal levels, and larger samples of
KMOS/MOSFIRE H$_{\alpha}$/H$_{\beta}$/[OII] emitters, just to mention few examples.
But still, astronomers must  rely on fitting the galaxy spectral energy distributions (SEDs) with galaxy models which imply,
by nature, some assumptions on several galaxy properties, including  their stellar initial mass function (IMF),
the chemical composition and its evolution, the dust-attenuation law and the star formation histories (SFHs).
Uncertainties in the underlying single stellar populations models, SSPs, also affect the
estimate of the energy output of the stellar populations,
mostly because of the contribution of stars of low and
intermediate mass experiencing the thermally pulsing asymptotic giant branch phase (TP-AGB) \citep{Maraston2006, Cassara2013,Villaume2015}. 
Several authors have addressed how  different assumptions can influence the SED-fitting results: for instance \citet{Papovich2001} and \citet{Conroy2009} 
have investigated the effect of the IMF, \citet{Conroy2009} have also studied the effect of metallicity evolution and \citet{Maraston2006}
the effect of stellar evolution models.
For a detailed review on these issues, we refer to \citet{Conroy2013}.

Notwithstanding these limitations, the comparison of the broadband spectral energy distribution of galaxies with that of a population
of stars with given IMF, SFH, age, dust reddening and metallicity remains one of the most widely used method to 
infer the physical properties of galaxies over a wide redshift range \citep[][just to mention a few]{Sawicki1998,Papovich2001,Guo2012}.
When modeling the SED of galaxies, some parametrization of the SFH must be used \citep{Ilbert2010,Lee2010,Wuyts2011,Ilbert2013}.
The SFH can in principle be arbitrarily complex, but simple forms are often adopted: one of the 
frequently used form is a declining exponential model, where SFH $\propto exp(-t/\tau)$, as
would be predicted from a closed box model of galaxy evolution \citep{Schmidt1959,Tinsley1980}.
It is worth recalling that the closed-box model of galaxy evolution, applied to study the chemical history of the solar neighborhood, 
fails to explain the metallicity distribution observed among old field stars, giving rise to the so called
G-dwarfs problem. Models with infall are able to solve this issue, since they avoid the excess of very low
metal stars: the metallicity increases faster, and few stars are formed at very low metallicities \citep{Chiosi1980,Bressan1994,Pipino2013}.
However, the use of exponentially declining SFH reproduces the optical/near IR colors of local spiral galaxies \citep{Bell2000}
and the global evolution in the SFR density for z<2 \citep{Nagamine2000}.

During the last decades, the advent of large redshift surveys have enabled the identification
and study of a large number of high redshift galaxies \citep{Davis2003,Steidel2003,Ouchi2008,Vanzella2009,Kashikawa2011, 
Bielby2013,LeFevre2005, LeFevre2013}, and contemporaneous 
studies from hydrodynamical simulations and semianalytic models \citep{Finlator2007,Finlator2011}
have suggested the need for rising star formation histories when studying the properties of galaxies at z>2.
\citet{Finkelstein2010} and \citet{Papovich2011} require rising star formation histories (at least on average)
to explain the evolution of the UV luminosity function.
The SFRs and stellar masses for galaxies at z >2 appear inconsistent with their having forming stars following
 an exponentially declining or constant star formation history prior to the epoch during which they are observed \citep{Reddy2012}.
\citet{Lee2009} analyzed the SEDs of high-z theoretical galaxies and 
concluded that the use of a single exponentially-decreasing SFH underestimates SFRs and
overestimates ages by a factor of $\sim$ 2 in both cases: subsequent works \citep{Lee2010,
Maraston2010, Wuyts2011} confirmed this result and concluded that models
with rising SFHs provided a better fit to high-z SEDs and produce SFRs in better agreements with 
other indicators.
Models with rising star formation histories in general lead to high SFRs, since their SED is always dominated by young stars
implying a narrower range of UV to optical fluxes, on average requiring a higher dust attenuation than for other SFHs \citep{Schaerer2005}.

There are different opinions regarding which functional form one should adopt for rising SFHs.
\citet{Maraston2010}, \citet{Pforr2012} advocated exponentially increasing SFHs, while \citet{Lee2010} suggested
the use of delayed $\tau$ models (SFH $\propto t \times exp(-t/\tau)$). The data do not favor one functional form over the other,
and the basic conclusion is that the model SFHs library must be sufficiently diverse to allow for a wide range in
SFH types \citep{Conroy2013}, especially when studying galaxies at high redshift, where a lot of uncertainties still hold.
One of the main arguments often invoked to support rising star formation histories is the small scatter in the SFR-M$_{*}$ relation \citep{Schaerer2013}.

Up to z $\sim 2$, the correlation between star formation rate and stellar mass of galaxies, and its evolution with redshift,
has been extensively studied in the last years by many
authors \citep[e.g.][]{Daddi2007,Elbaz2007,Rodighiero2011}: the galaxies following this relation define 
what has been called the main sequence of star forming galaxies \citep{Noeske2007}.\\
The slope and the scatter of this relation together with its evolution in redshift put constraints on the SFHs of galaxies as a function of their mass \citep{Buat2012}.
It is worth underlining that the SFR is generally derived from observables (UV and IR luminosity) which depend
on timescales t $\gtrsim$ 100 Myr and which are assumed to be in an equilibrium value, which is only reached after this timescale
and for constant SFR.
The consequence of these assumptions entering the SFR(UV) or SFR(IR) calibrations \citep{Kennicutt1998}  is that the observational scatter is smaller than the 
true scatter in the current SFR when typical ages are less than 100 Myr or for shorter timescales, which is especially true at high redshift \citep{Schaerer2013}.
\citet{Schaerer2013} and \citet{DeBarros2014} showed  that the idea of a simple, well-defined 
star forming sequence with the majority of star forming galaxies showing a tight relation between
stellar mass and SFR suggested by studies at low redshift \citep[z $\leq$2,][]{Daddi2007, Elbaz2007, Noeske2007} may
not be appropriate at high redshift. A relatively small scatter is only obtained assuming star formation 
histories constant over long timescales (t $\gtrsim$ 50 Myr), while a significant scatter is obtained for models 
assuming rising or delayed star formation histories which are often suggested in recent works. 

\citet{Tasca2015} recently presented a study on the evolution with redshift of the SFR-M$_{*}$ relation and of the sSFR of a sample
of 4531 galaxies from the VUDS survey, with spectroscopic redshifts between 2 and 6.5.
The values of M$_{*}$ and SFR have been obtained by fitting all the available multi-wavelenghts data
with Le Phare code \citep{Arnouts1999,Ilbert2006} and using a range of templates coming from the \citet{Bruzual2003} models.
The assumed IMF is the Chabrier IMF, while the adopted SFHs are exponentially declining ($SFR \propto e^{-t/\tau}$)
and two delayed SFHs models with peaks at 1 and 3 Gyrs.
They conclude that the $logSFR-logM$ relation for star forming galaxies (SFGs) remains linear up to z=5 but the SFR
increases at a fixed mass with increasing redshift. 
For stellar masses M$_{*}$ $\geq$ 10$^{10}$ M$_{\odot}$ the SFR increases by a factor 1.7 from redshift
z $\sim$ 2.3 up to z $\sim$ 4.8.

In this paper we aim at measuring the stellar mass and the SFR using
different assumptions about star formation histories in order to investigate the nature of 
the scatter of SFR-M$_{*}$ relation.
We use a sample of 2995 VUDS galaxies with reliable spectroscopic 
redshift $z_{\rm spec}>2$ \citep{LeFevre2015}.
The SED fitting is performed using both the photometric data
and the spectra and a new set of galaxy models
built from  SSPs recently developed by
\citet{Cassara2013} and an ample choice of SFHs, 
described in Appendix ~\ref{app:SSPs} and ~\ref{app:galaxies}.
The treatment of the dust absorption and
re-emission is based on the radiative transfer model.
The plan of the paper is the following: 
in Section 2 we describe the properties of the galaxy sample and in Section 3 the galaxy models and the allowed range of free parameters.
In Section 4 we show the results of
a test on a set of mock galaxies, in order to validate the use of the galaxy models.
In Section 5 we present the SED fitting tool and procedure, while in Section 6 we present the 
results which are discussed in Section 7. Section 8 is devoted to a short summary.
We assume a cosmology with $\Omega{_m}$=0.27, $\Omega{_\Lambda}$=0.73 and 
$H_{\rm 0}$=70.5 km s$^{-1}$ Mpc$^{-1}$.

\section{The VUDS sample}

The VIMOS Ultra Deep Survey \citep[VUDS,][]{LeFevre2015} is a spectroscopic redshift survey devised to study
galaxy evolution in the high redshift Universe, taking advantage of the multiplexing capabilities
of VIMOS on VLT. 
This analysis is based on a subsample of the 7,843 galaxy redshifts measured in three separate fields (COSMOS, VVDS-02h and ECDFS) where ample photometric data are also available.

As reported in \citet{Tasca2015} each of the fields has multi-band photometry covering at least from broad band $u$ to Spitzer-IRAC 4.5 $\mu$m. 
The COSMOS field  has the most extensive photometric set composed by more than 30 bands including standard broad band and medium band photometry \citep{Laigle2016}. 
The broad band photometric databases of VVDS-02h and ECDFS have been presented respectively in   \citet{LeFevre2015} and \citet{Cardamone2010}.
VUDS targets have mainly been selected  on the basis of their photometric redshifts
requiring that either the first or second peak in the photometric redshift 
probability distribution satisfy the condition 
$z_{\rm phot}+1\sigma \geq 2.4$ and $i_{\rm AB} \leq 25$.
Exposure times were 14h  with both the VIMOS Blue and Red Low Resolution grisms with slits one arcsecond wide.
The blue and the red parts of the spectra were normalized in the
common wavelength range and joined together to obtain a spectrum covering from 
3650 to 9350 $\AA$.\\
Redshifts were measured independently by two astronomers and disagreements were discussed before assigning them
a confidence level.

For this work we selected all objects with $z_{\rm spec} \geq 2$ and a confidence level >75\%, i.e.
with flags 2,3,4 and 9 \citep[see][for more details about the redshift measurement and the survey
in general]{LeFevre2015}.
3948 VUDS galaxies (1994 in the COSMOS field, 1499 in the VVDS-02h and 455 in the ECDFS field)
satisfy these requirements.
The SED fitting was performed considering all the available broad and medium band photometric data.

\begin{figure}[!ht]
   \centering
  \includegraphics[width=0.50\textwidth]{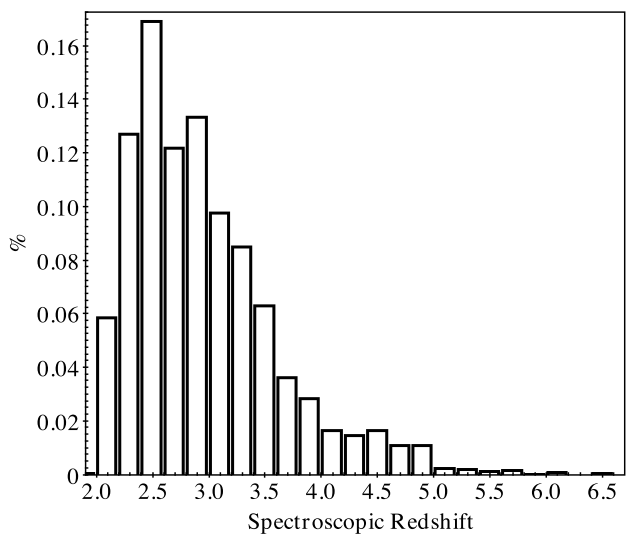}
     \caption{The normalized redshift distribution of VUDS galaxies used in this study.}
        \label{ist_redshift}
\end{figure}

Fig.~\ref{ist_redshift} shows their redshift distribution:  61\% of the galaxies have 2<z$_{\rm spec}$<3,
31\% have 3<z$_{\rm spec}$<4, while only 8\% have z$_{\rm spec}$>4. 
The redshift range is 2.00<z$_{\rm spec}$<6.54, while the mean and median value are 
respectively z$_{\rm spec,mean}$=2.96 and z$_{\rm spec,median}$=2.83.

\section{Galaxy models}\label{models}

\begin{table*}
\caption{Input parameters for SED fitting with GOSSIP+.}
\centering
\begin{tabular}{c c c}
\hline
SSPs &  Range \\
\hline
Metallicities & 0.0004, 0.004, 0.008, 0.02, 0.05  \\
Ages & 0.03 - 3 Gyr \\
Optical Depths in the $V$ band $\tau_{\rm V}$ & 0.01, 0.03, 0.05, 0.08, 0.1, 0.3, 0.5, 0.8, 1, 1.3, 1.5, 1.8, 2, 2.3, 2.5, 2.8, 3, 3.3, 3.5, 4\\
\hline
IMF & Range \\
\hline
Salpeter law &   0.1-100 M$_{\odot}$ \\
$\zeta$   & 0.35 - 0.39 - 0.50 \\
 slope & -2.35 \\
\hline
SFHs &  Range \\
\hline
infall time $\tau_{\rm infall}$ & 5 - 0.30 Gyr\\
efficiency of the SF $\nu$ & 14 - 0.35\\
exponent of the Schmidt function $k$  & 1\\
\hline
IGM transmission & 19\% to 100\% at z$_{\rm spec}$=3.0 \\
 (seven possibilities at any z$_{\rm spec}$)                & 5\% to 50\% at z$_{\rm spec}$=5.0 \\
\hline
\label{table_models}
\end{tabular}
\end{table*}

\begin{figure}
   \centering
   \includegraphics[width=0.5\textwidth]{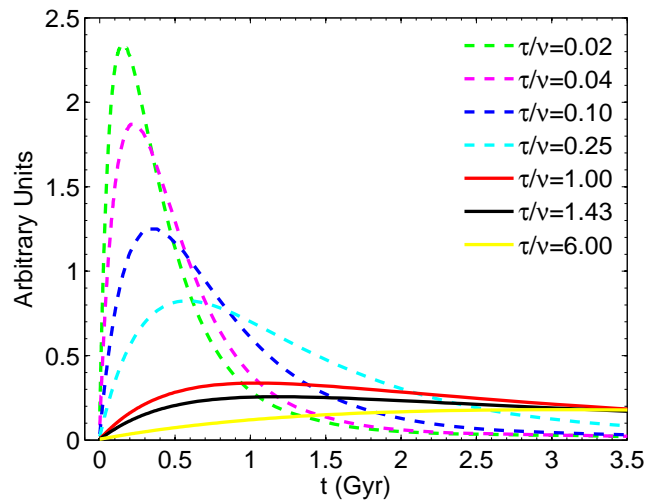}
      \caption{The evolution with time of the seven main families of the SFHs used in this work.
			Dashed and solid lines represent, respectively, SFHs typical for bulge-like models
			and disk-like models.}
         \label{SFHs}
   \end{figure}

The SED fitting is performed using a large database of dusty model galaxies (about 30,000 evolutionary models)
obtained varying the physical input parameters. 
The galactic models have been calculated starting from   
the extended library of isochrones and SSPs of different chemical compositions and ages.
 The definition of age throughout the paper means the time passed since the first stars started forming in a galaxy. In the models we take into
account a state-of-the-art treatment of the TP-AGB phase and including the effect of circumstellar dust shells
around AGB stars \citep{Cassara2013}.
Even if the AGB phase is short lived, the AGB stars are very bright,  can reach very low effective
temperatures and  easily get enshrouded in a shell of self-produced dust, which reprocesses
the radiation from the underneath stars.
AGB stars have a non negligible impact on the rest-frame near IR spectra of galaxies
and they can significantly alter the M$_{*}$/L ratio of intermediate-age populations.
The contribution of the TP-AGB stars is limited to galaxies dominated by stars with ages in the range 0.3-2 Gyr,
depending on metallicity \citep{Maraston2005,Bruzual2007,Marigo2008}. 
At z $\sim$ 2.5, a sizable fraction of the stellar populations has mean ages in the range 0.5 - 1.5 Gyr,
thus mostly affected by the TP-AGB stars \citep{Zibetti2012}, and even accounting for the high uncertainty 
affecting the theoretical modeling of this phase, stellar population
models including the TP-AGB stars allow for a better determination of galaxy ages
and hence stellar masses, fundamental quantities for studying galaxy formation and
evolution \citep{Maraston2006}. 
From these brief considerations, the correct inclusion of a more realistic modeling
of the TP-AGB phase in SSPs and hence galaxy models appears not negligible, both for studies in the local Universe and 
at high redshifts.\\
A physically realistic coupling between the populations of stars 
and the effect of attenuation and emission by dust 
is  critical to determine the properties of galaxies: 
see Appendix~\ref{app:SSPs} for more details about the SSPs and for the treatment of dust extinction and re-emission,
and Appendix~\ref{app:galaxies} for the features of the Composite Stellar Populations (CSPs).
The models do not include the contribution of the emission lines: their net effect in the
derivation of both stellar masses and SFRs has been explored by many authors: in our redshift range, the SED-derived masses could change
by 0.1-0.2 dex \citep{Ilbert2009,DeBarros2014,Salmon2014,Tasca2015}, even though according to \citet{Salmon2014}
this may affect about 65\% of the galaxies.
As for the star formation rate, we use the Schmidt law \citep{Schmidt1959} which, in the 
formalism of adopted chemical evolution models (see Appendix~\ref{app:SSPs}), becomes:
\begin{equation}
SFR(t)= \nu M_{g}(t)^{k}
\end{equation}

\noindent where M$_{\rm g}$ is the mass of the gas at the time $t$, $k$ yields the dependence of the star formation rate on the gas content
while the factor $\nu$ measures the efficiency of the star formation process.
In this type of models, because of the competition between the gas infall, gas consumption by star formation, and gas ejection
by dying stars, the SFR starts very low, grows to a maximum and then declines.
The functional form that could mimic the trend for the gas infall prescription is a delayed exponentially declining law $t \times exp(-t/\tau)$.
The time scale $\tau_{\rm infall}$ roughly corresponds
to the age at which the star formation activity reaches the peak value (see Appendix~\ref{app:SSPs}).
The shape of star formation rate is mostly driven by the two aforementioned parameters $\tau_{\rm infall}$ and $\nu$.
The final interplay between them drives its evolution with time (see Fig.~\ref{SFHs}). 
The complete formalism of the chemical evolution models can be found in its
original form in \citet{Tantalo1996} and \citet{Portinari2000} while a short summary is provided in
\citet{Cassara2015}.  
With this law for the star formation, we are able to model two main types of objects:
the first type of models are what we call ''bulge-like'' models, characterized by high values of $\nu$ and low values of $\tau_{\rm infall}$, 
with a rapid rise of the star formation rates, a peak reached 
in a relatively short timescale (on average 0.5 Gyr) and a declining phase.
These models reproduce the chemical pattern in the
gas of elliptical galaxies at both low \citep{Piovan2006b,Pipino2011} and high redshift 
\citep[e.g.][]{Matteucci2002,Pipino2011}.
The second type of models are what we call ''disk-like'' models, characterized by low values of $\nu$, together with high values of $\tau_{\rm infall}$,
which show a slow rising and declining SFHs: they reproduce disk galaxies in the
local Universe \citep{Piovan2006b,Pipino2013}.\\
Many prescriptions for the SFHs can be found in the  literature: our choice of using a variety of SFHs and not the 
classical $\propto$ e$^{(t/\tau)}$ rises from the following reasons.
As amply described in Appendix~\ref{app:galaxies}, model galaxies are usually obtained considering a convolution of 
SSPs of different age and metallicity, weighted by the SFHs: in our case, 
we also considered the chemical enrichment which is described by the infall model \citep{Chiosi1980,Pipino2013}.
In this framework, the Schmidt function is a \textit{physical} prescription which relates the gas content to the star formation activity,
allowing to model different types of objects by varying its parameters \citep{Buzzoni2002}.\\
The original catalog of 28 different SFHs (see Table~\ref{table_galaxies}) can be grouped in 
7  families shown in Fig.~\ref{SFHs}, according to the increasing ratio between $\tau_{\rm infall}$ 
and the efficiency of the star formation rate $\nu$.
This ratio is in turn representative of the two populations of model galaxies, bulge-like (dashed lines in Fig.~\ref{SFHs}) and disk-like
(solid line in Fig.~\ref{SFHs}). 	
Table~\ref{table_models} gives the range of free parameters for the dusty SSPs, the galaxy models, the SFHs and the IGM transmission.
We recall that each combination of $\tau_{\rm infall}$, $\nu$ and $\zeta$ (Table~\ref{table_galaxies}) gives rise 
to 28 sets of evolutionary galaxy models. $\zeta$ describes the fraction of total mass in form of
stars stored in the IMF above a given mass M$_{*}$, which is the minimum mass contributing
to the nucleo-synthetic enrichment of the ISM over a timescale of the order of the
galaxy life \citep{Bressan1994,Tantalo1996,Cassara2015}.
This is equivalent to fixing the lower limit of the integral on the SSP mass used to normalize the IMF. 
While the upper limit of the integral  could be fixed to  100 or 120 M$_{\odot}$ because 
massive stars are a small fraction of a stellar population defined with a Salpeter IMF, 
the lower mass stars  have a higher contribution to the SSP global mass but not to its luminosity. 
Therefore, the adoption of a smaller limit for the IMF (e.g. 0.01 M$_{\odot}$) instead of  0.1 M$_{\odot}$
can lead to a larger number of low mass stars and  create a stellar population less luminous (in mass unity) and with less capability of enriching the ISM (see \citet{Bressan1994, Tantalo1996, Portinari1999, Piovan2011a,Cassara2012,Cassara2015} for an extensive discussion on this topic).
Following the suggestion of \citet{Tantalo1996,Portinari2004, Cassara2012} we adopt three values for $\zeta$ for our models, namely 0.35, 0.39 and 0.50.\\
The galaxy models have been calculated at varying the possible values of the optical depth
$\tau_{\rm V}$, and hence of the database of dusty SSPs. It is worth underlining the relation between
$\tau_{\rm V}$ and A$_{\rm v}$: A$_{\rm v}$=1.086 $\times$ $\tau_{\rm V}$
(more details in Appendix~\ref{app:dust}).

\section{Validation of the models: test with mock galaxies}

\begin{figure*}
\subfigure{
 \includegraphics[width=0.50\textwidth]{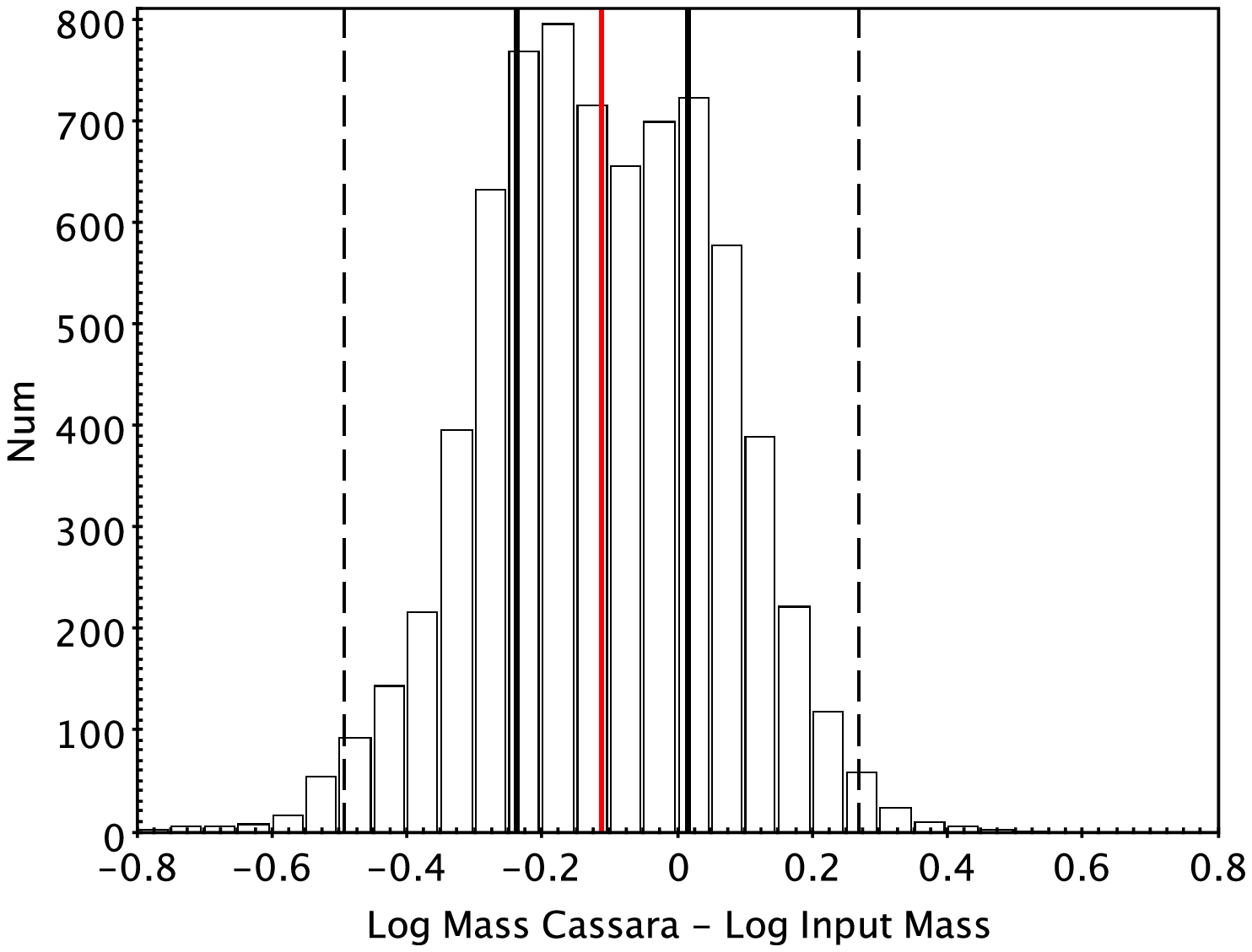}}
 \subfigure{
 \includegraphics[width=0.50\textwidth]{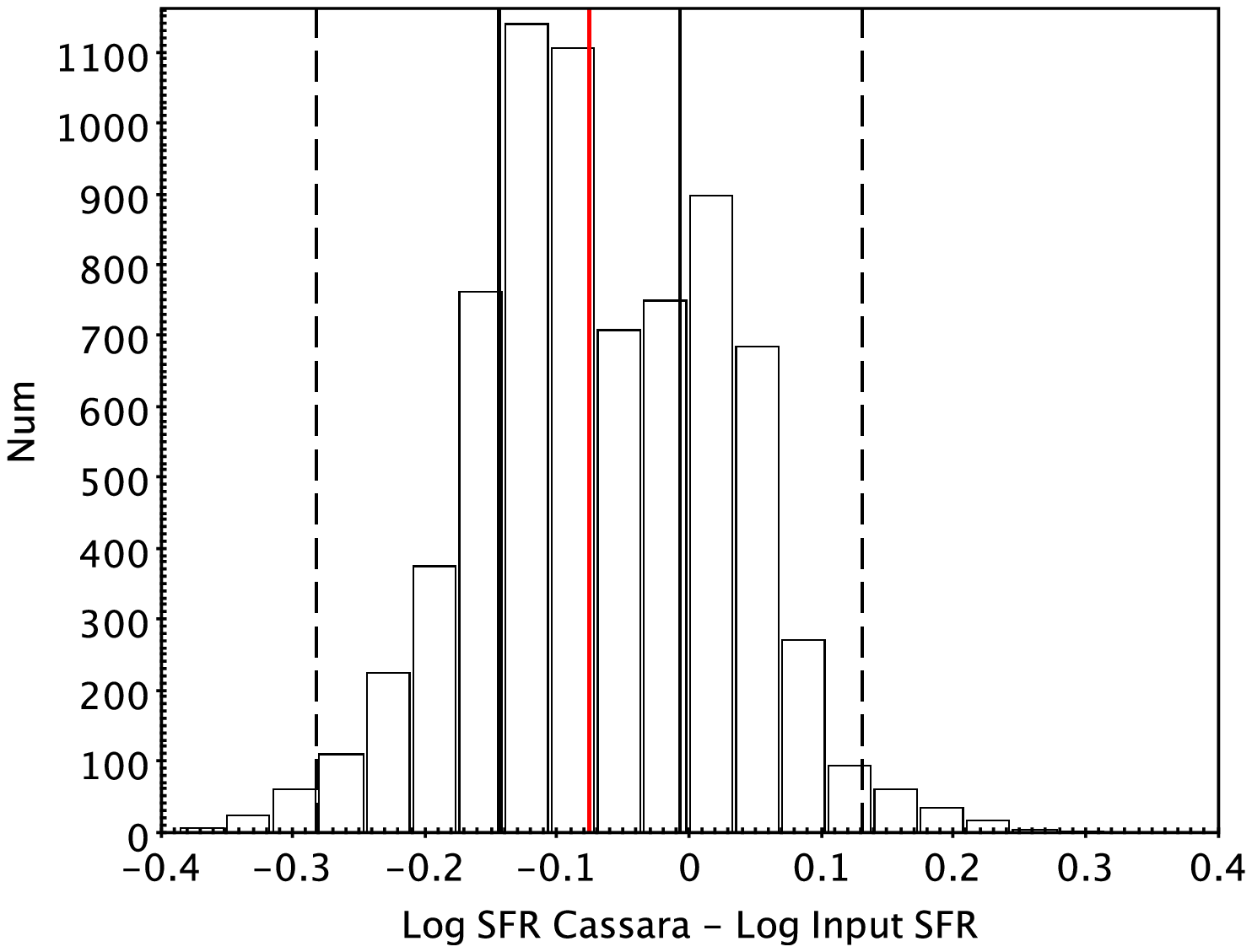}}
     \caption{Left panel: the distribution of the differences between the values of stellar mass inferred from SED fitting on the mock galaxies and the 
input values. The red solid line is the median. Black dashed and solid lines indicate, respectively, 3 and 1 times the median absolute deviation (MAD).
Right panel: as in the left panel, but for the values of SFRs.}
        \label{mock}
\end{figure*}

We performed a test on a set of mock galaxies in order to 
inspect the capability of our models to recover the input values of stellar masses and SFRs.
The mock galaxies are based on galaxy models from the PEGASE library \citep{Fioc1997}.
The use of this library is justified by the self-consistent treatment of the evolution of 
galaxy physical properties, without the need of fixing and choosing, e.g., the metallicity and the extinction of the populations,
as in \citet{Bruzual2003} models. These models do not adopt the same SSPs  used in this paper and
have a different approach for the dust extinction.
The IMF for the mock sample is the Salpeter law integrated between two fixed values \citep{Fioc1997}.
These models are characterized by ages ranging from 0.1 to 2.5 Gyr and timescales of star formation histories between 0.1 and 7 Gyr. 
These choices lead to models with star formation histories which can be exponentially declining, constant or delayed exponential, 
depending on the combinations of their age and timescale.  
The final sample of synthetic models comprise 7350 galaxies, with stellar masses between 10$^{8}$ and 10$^{12}$ M$_{\odot}$ and SFRs between 
1 and 200 M$_{\odot}$/yr, following the main sequence at z=2 as defined by \citet{Daddi2007}, with a slope of $\alpha$=0.87.
The SED of the mock galaxies have photometry from the $u$ band to the IRAC 4.5 $\mu$m band (the longer wavelength that can be simulated 
using this library to create mock galaxies) and their simulated spectra
cover a spectral range from 3500 to 9600 $\AA$ with an average S/N $\sim$ 3.5.
VUDS galaxies reach a S/N = 5 on the continuum at 8500 $\AA$  for i$_{\rm AB}$ = 25 \citep{LeFevre2015}.
The photometric data come with random errors, while spectra have been computed adding  realistic noise spectra.\\
We use GOSSIP+ (see later) to fit the mock galaxy SEDs with our models, and the results of this exercise are shown in the panels of Fig.~\ref{mock}.
Concerning the stellar masses, the distribution of the differences between the values inferred from the SED fitting using the set of models presented in this paper
and the input values of PEGASE mock galaxies is shown in the left panel of Fig.~\ref{mock}.
The median value of the differences is -0.11 dex (red line in the left panel of Fig.~\ref{mock}) with 1 median absolute deviation (MAD) of 0.12 dex
(black solid lines in the same panel). The dotted lines show 3 times the MAD.
Stellar mass is considered as the most reliable parameter estimated by SED fitting, since relatively
small differences are found while varying the assumptions for the star formation histories and/or 
the dust extinction.
\citet{Finlator2007} estimates that differences due to different assumptions on SFHs are around 0.30 dex and
\citet{Yabe2009}, adding effects of metallicity and extinction law, estimate differences not higher than 0.60 dex.\\
The distribution of the differences between the SFRs as inferred from GOSSIP+ with respect to input SFRs presents a median value of -0.07 dex 
(red line in the right panel of Fig.~\ref{mock}) with 1-MAD of 0.06 dex (black solid lines in the same panel).
The dotted lines indicate the 3-MAD dispersion.\\
We notice a slight bimodality in both panels of Fig.~\ref{mock}.
A significant number of galaxies presents differences in SFRs between -0.07 dex and -0.13 dex (left peak in the histogram, right panel of Fig.~\ref{mock}),
 while a smaller percentage presents differences in SFRs between 0 and 0.05 dex  (right peak in the same histogram).
This feature appears also in the left panel of Fig.~\ref{mock}, though not so strong as in the aforementioned one.
The driver of this bimodality is the adoption of different values of $\zeta$ with respect to the single value allowed for the mock galaxies.
This bimodality is not observed in any of the following plots presenting the results of the SED fitting on the VUDS galaxies. None of the physical properties shows any correlations with the values of $\zeta$. \\
These differences in stellar mass and SFR estimates  are fully acceptable, taking into account  the uncertainties coming from the adoption of the SED fitting technique \citep{Conroy2013}.
From the tests we performed and the above discussion, we can conclude that 
the template library used features a sufficiently broad range of SFHs and extinctions to encompass those in the PEGASE library, and that the SED fitting works properly.
We can confidently use
the galaxy models to recover the physical properties of the VUDS galaxies.

\section{SED fitting tool and procedure} \label{tool}

\begin{figure}
   \centering
   \includegraphics[width=0.50\textwidth]{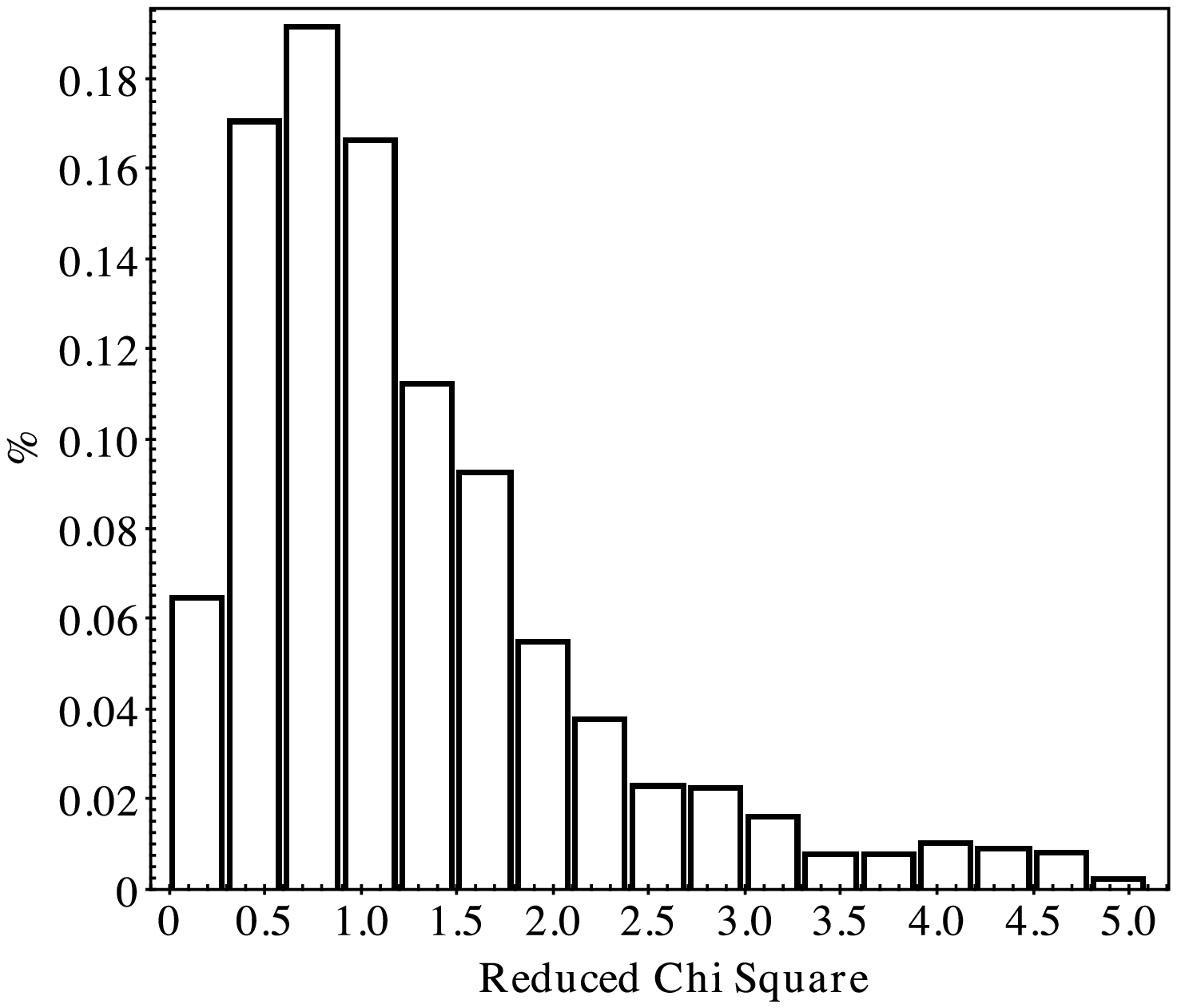}
      \caption{The normalized distribution of the objects sample with $\chi^2$<5.
              }
         \label{isto_chi}
   \end{figure}
   
   \begin{figure*}
\centering
{\includegraphics[width=0.75\textwidth]{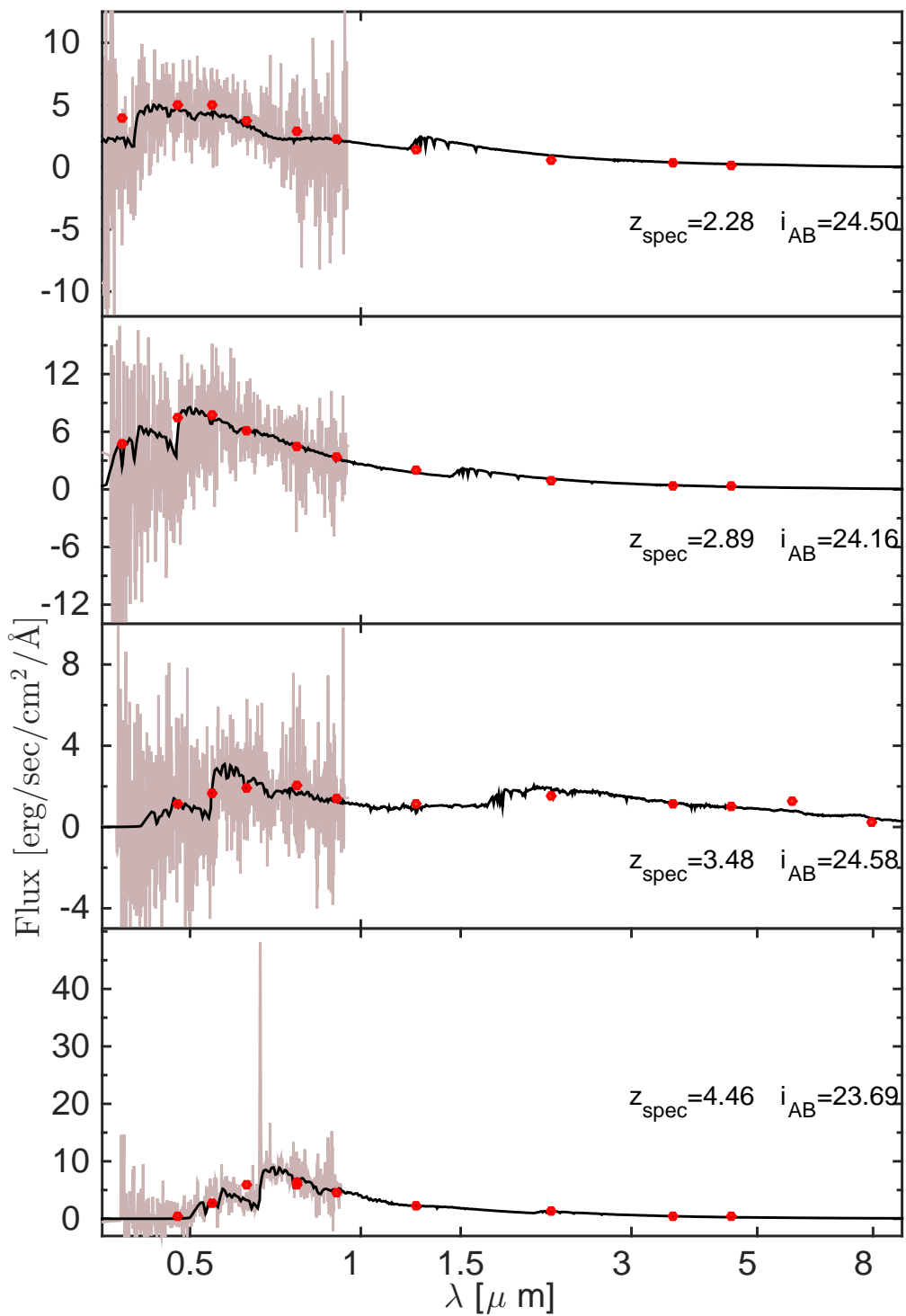}}
\caption{Examples of model fits to the data using GOSSIP+. The dark curve represents the best-fitting model,
the grey curve the spectra, the red points are the photometric determinations.
Fluxes are given in units of 10$^{19}$ [erg/sec/cm$^{2}$/$\AA$].
The redshifts and the i$_{\rm AB}$ of the objects are shown in each panel. 
}
\label{SED}
\end{figure*}
 	
We make use of GOSSIP+ (Galaxy Observed-Simulated SED
Interactive Program), a spectral/SED fitting software.
It has been developed by the PANDORA group at INAF-IASF Milano
\citep{Franzetti2008} and partially modified and completed
at LAM in Marseille.  The full description of the new version
of the sofware is presented in  \citet{Thomas2016}: for the sake of clarity, here we make  a short summary of its capabilities.
GOSSIP+ is a software that is able to combine the
spectroscopic information with the photometric data points. 
One major improvement implemented in GOSSIP+ deals with the IGM attenuation.
Most current SED fitting softwares use the Madau's prescription \citep{Madau1995}. 
This means that for a given redshift a single extinction curve is
considered, while GOSSIP+ allows to choose among seven IGM extinction curves
at any redshift. In order to account for the vastly different number 
of spectral data points with respect to the photometric data points, a combined $\chi^2$ is computed by GOSSIP+ as the sum of the
reduced $\chi^2$ of the photometric and of the spectrum data fit, in order to
evaluate the two different data sets with the same weight.
Being the sum of two reduced $\chi^2$ we adopt as a definition of a total reduced $\chi^{2}$ the following:
$\chi_{\rm red}^{2}=0.5  (\chi_{\rm phot}^{2}+ \chi_{\rm spec}^{2}$).
As for the prominent emission lines in the spectra, GOSSIP+ gives the possibility 
to define some avoidance regions of the spectra (which are then redshifted 
to each object redshift) which will be avoided in the fitting procedure.\\
Stellar masses, SFRs, dust absorption, age and metallicity of the stellar 
populations, UV luminosity and level of IGM extinction are obtained 
simultaneously employing a $\chi^2$ minimization to find the best-fit model
considering the age of the Universe at the observed redshift as an upper limit
for the choice of the models. 
The agreement between spectroscopy and photometry is crucial to produce a good fit. As discussed in \citet{Thomas2016}, it can happen that there is a mismatch between these two datasets and this affects the quality of fit. In \citet{Thomas2016} a visual inspection was performed in order to define the quality of the fit. We decided to select the final sample of galaxies in the following way: as a first step, we excluded galaxies with values of the total reduced $\chi_{\rm red}^{2}$ >5, and as a second step, we excluded galaxies with values of reduced photometric $\chi_{\rm phot}^{2}$>5. Finally, a visual inspection of the excluded galaxies confirmed that in this way we retain only galaxies with a reliable fit.
High values of $\chi^2$  are almost always due to a very reduced set of photometric data and/or a spectrum 
with low S/N, and thus the physical properties that can be inferred could be highly uncertain.
The $\chi_{\rm red}^{2}$ distribution of the 2995 galaxies is shown in Fig.~\ref{isto_chi}, while Fig.~\ref{SED} shows the results of the fitting procedure for four galaxies at different redshifts.
The final sample comprises 2995 galaxies (1737 in the COSMOS field, 1005 in the VVDS-02h and 253 in the ECDFS field) with reliable spectroscopic redshift between
2.00 and 6.54. 

\begin{figure}
   \centering
   \includegraphics[width=0.50\textwidth]{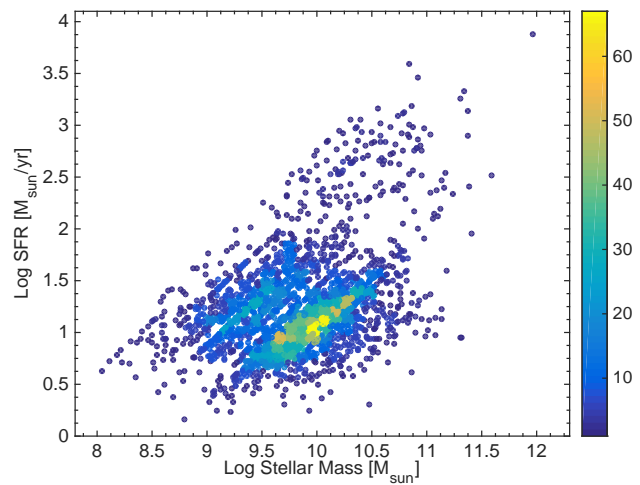}
      \caption{SFR-M$_{*}$ density map for the complete sample of VUDS galaxies.}
         \label{density}
   \end{figure}

\section{SED fitting results}\label{results}

\begin{figure*}
   \centering
   \includegraphics[width=1\textwidth, clip]{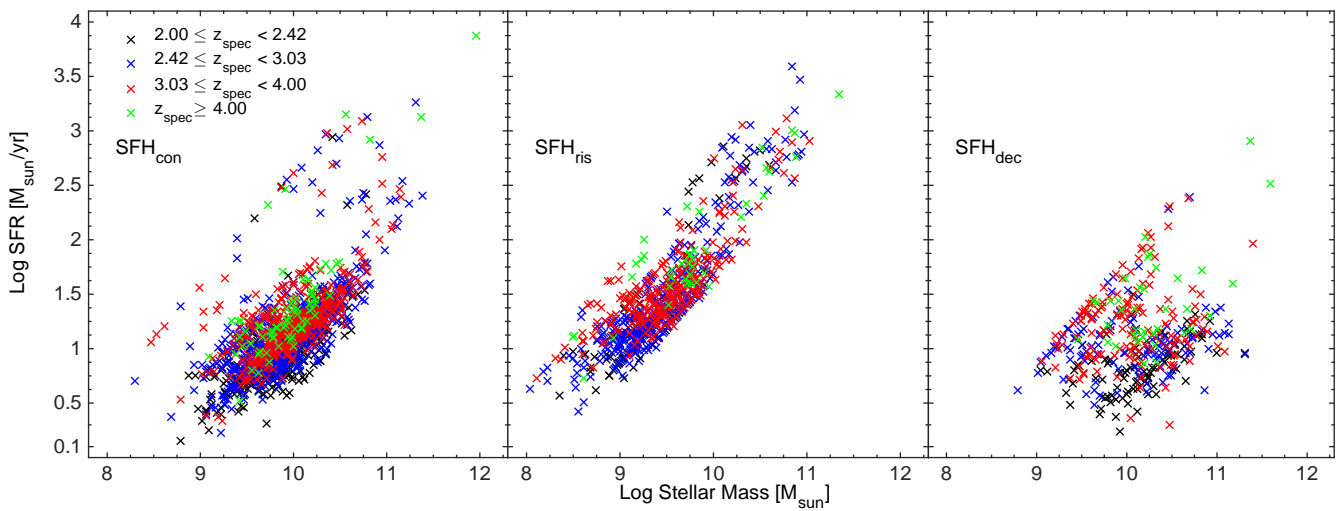}
     \caption{SFR-M$_{*}$ plane: galaxies have been color coded according to their redshift range as the legend indicates.
  The left panel presents  galaxies best fit by SFH$_{con}$ models, in the central panel galaxies best fit by SFH$_{ris}$ models and the right panel            galaxies best fit by SFH$_{dec}$ models (see text for more details about these definitions). }  
\label{picco}
   \end{figure*}

In this paper we want to investigate what happens to the SFR-M$_{*}$  relation when 
the SED fitting is performed with a set of models with different SFHs. 
Fig.~\ref{density} shows the outcome of our SED fitting procedure on 2995 VUDS galaxies spanning a wide range of redshifts.
The spread in SFRs for a given stellar mass is quite important (of the order of 1 
dex) and there is a suggestion of two distinct "main sequences" showing SFRs
differing by 0.6 dex, plus a population of highly star forming galaxies 
amounting to 5\% of the sample.

\begin{figure}
   \centering
   \includegraphics[width=0.50\textwidth]{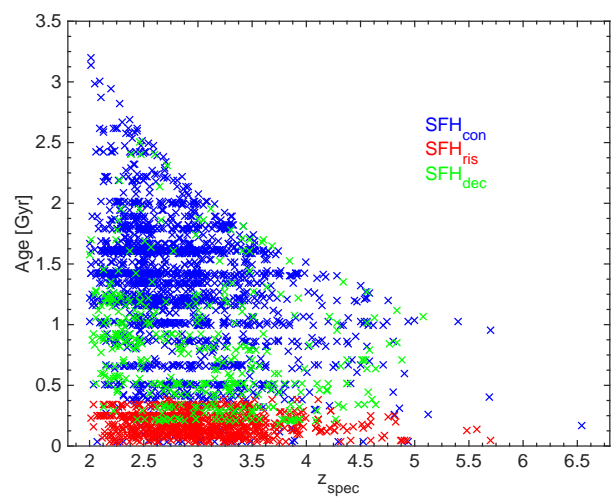}
      \caption{The inferred values of ages as a function of redshift. Galaxies have been color coded according to their best fit SFH.}   
       \label{age_z}
   \end{figure}

The spread in SFRs could be due either to the wide redshift range covered by the VUDS galaxies, because
of the evolution of the SFR-M$_{*}$ relation between z $\sim$ 5 and z $\sim$ 2
\citep{Elbaz2007,Fumagalli2014,Tasca2015} or to the use of different star formation histories \citep{Salmon2014}.
In order to  disentangle these two effects we divide our sample both in redshift bins and according to the type of SFHs  
of the best-fit model. 
As far as redshift concerns, we subdivide the sample in three redshift bins of approximately 570 Myr (2.00 $\leq$ z$_{\rm spec}$<2.42, 2.42 $\leq$ z$_{\rm spec}$<3.03,  3.03 $\leq$ z$_{\rm spec}$<4.00)  and the fourth one including all galaxies with z$_{\rm spec}$ $\geq$ 4.00.
As for the SFHs, we consider three different families: a) SFH$_{con}$ 
characterized by low star formation efficiency $\nu$ and high values of the infall
time scale $\tau_{\rm infall}$ (shown with solid curves in Fig.~\ref{SFHs}); b) SFH$_{ris}$  characterized by high star formation 
efficiency and low gas infall time scales (dashed lines in Fig.~\ref{SFHs}, rising part); c) SFH$_{decl}$: again, they present  high values for the star formation efficiency and low gas infall time scales (dashed lines in Fig.~\ref{SFHs}, declining part).

\begin{figure*}
\centerline{
\includegraphics[width=0.7\textwidth]{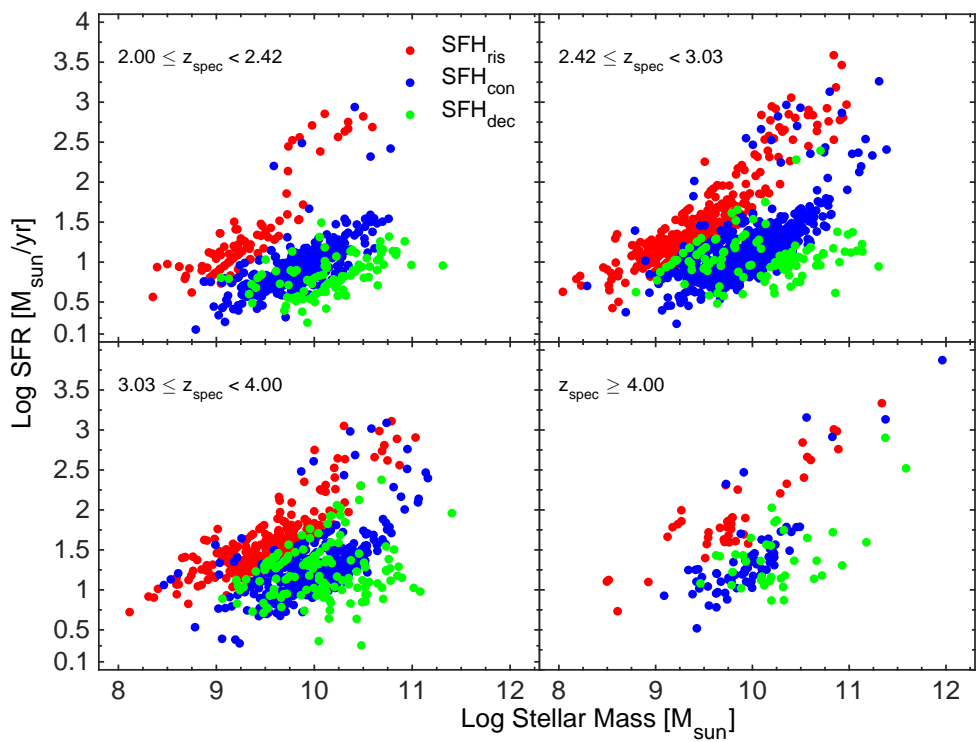}}
\caption{SFR-M$_{*}$ plane for galaxies in the different redshift range as the legend in each panel indicates.
Galaxies have been color coded according to their best-fit SFH.
In blue, galaxies best fit by SFH$_{con}$ models, in red galaxies best fit by SFH$_{ris}$ models and in green galaxies best fit by SFH$_{dec}$ models 
(see text for more details about these definitions).}
\label{all}
\end{figure*}

The three panels of Fig.~\ref{picco} show the SFR-M$_{*}$ plane for the three groups of SFHs, where 
galaxies have been color-coded according to the redshift bins. 
It is worth underlining that each panel is populated by galaxies of different redshift range. 
The left panel shows that a
classical "main sequence" arises from  galaxies fit by models with SFH$_{con}$ 
while galaxies best fit by models with rising SFHs SFH$_{ris}$ define 
a second "main sequence" with higher sSFR (Fig.~\ref{picco}, central panel) and, on average, younger ages.  Finally,
the right panel of Fig.~\ref{picco} shows galaxies fit by models with declining SFHs
SFH$_{dec}$. 
This last group of galaxies shows higher masses
than the others, moderate SFRs and older ages (see Fig.~\ref{age_z}). No clear main sequence is 
visible in this last group.

\begin{figure}
   \centering
   \includegraphics[width=0.50\textwidth]{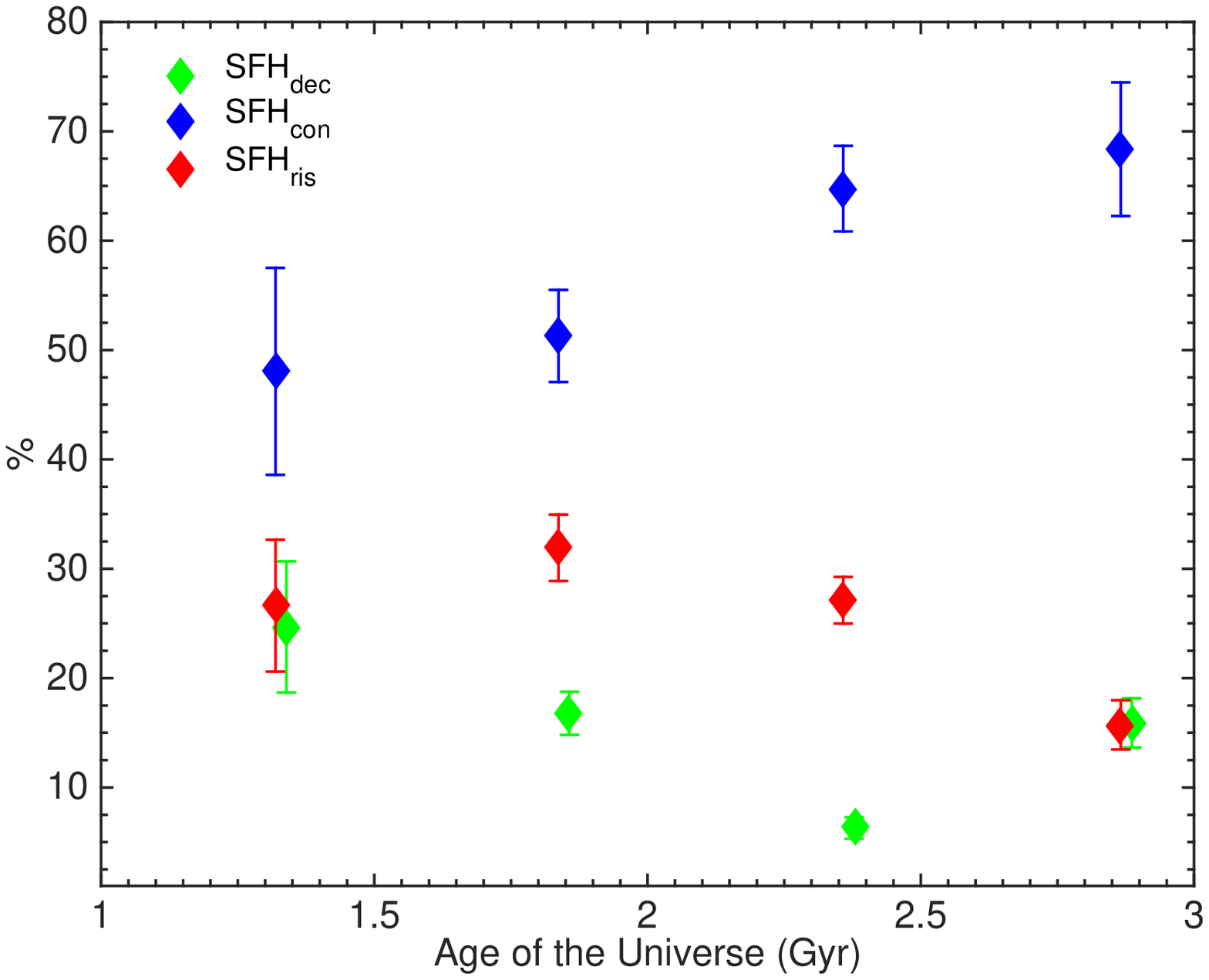}
      \caption{The variation of the percentage of galaxies fitted by different SFHs with the age of the Universe.
      The symbols are color coded according to the SFH. The green points have been slightly
 shifted to avoid overlap with the red ones.}   
       \label{rapporti}
   \end{figure}

\begin{table*}
\caption{Redshift evolution for galaxies with different star formation histories.}
\centering
\begin{tabular}{c c c c c}
\hline
SFH / $\Delta$ z$_{\rm spec}$    &  2.00 $\leq$ z$_{\rm spec}$ <2.42 & 2.42 $\leq$ z$_{\rm spec}$<3.03 & 3.03 $\leq$ z$_{\rm spec}$<4.00 & z$_{\rm spec}$ $\geq$ 4.00\\
\hline
 SFH$_{con}$ & 417 (68\%) & 893 (65\%) & 437 (52\%) & 75 (49\%) \\
 SFH$_{ris}$ & 96 (16\%) & 351 (27\%) & 272 (32\%) & 41 (26\%) \\
 SFH$_{dec}$ & 97 (16\%) & 112 (8\%) & 143 (16\%) & 38 (25\%) \\
\hline
\label{table_SFH}
\end{tabular}
\end{table*}

The panels of Fig.~\ref{all}  show
the SFR-M$_{*}$ plane for galaxies divided in the four redshift bins.
It is clearly visibile that all redshift bins show a similar scatter due to the presence of galaxies fitted by models with different SFHs.
From Fig.~\ref{picco} and Fig.~\ref{all} , 
it is evident that redshift is not the 
dominant factor for the spread observed in the SFR-M$_{*}$ plane.

We further note that the
proportion of galaxies fit by the models with different SFHs changes with the
age of the Universe. Galaxies occupying the classical main sequence (SFH$_{con}$, blue points) increase
from about 50\% to 68\% as the Universe gets older, while the more rapidly star
forming galaxies (SFH$_{ris}$, red points) decrease from about 30 to 15-20\%, as shown in Fig.~\ref{rapporti} and Table~\ref{table_SFH}.
The last point we would like to note is that those galaxies showing the highest
SFRs (greater than 100 M$_{\odot}$/yr) and representing 5\% of the sample are the
ones with higher optical depths.

\section{Discussion and Conclusions}

When used to fit the SEDs of high redshift star forming galaxies, our set of 
models including SFHs of different shapes produces a SFR-M$_{*}$ relation showing
a large spread. We have shown that such large spread is not due so much to the large redshift range, but to the different SFHs used.
This implies that the dominant stellar populations of galaxies of 
similar mass can have substantially different ages and  also quite different 
sSFRs. One could interpret the bimodal appearance of the SFR-M$_{*}$  relation as just due to the
limited number of models we have used for the SED fitting, i.e.
the discreteness of the combination of the $\nu$ and $\tau_{\rm infall}$ parameters.
A more extended model library with less discrete parametrization of $\nu$ and $\tau_{\rm infall}$ could, in principle, show
a more continuous distribution of galaxies in the SFR-M$_{*}$ plane.
 Nonetheless, some evidences a bimodal SFR-M$_{*}$ relation has been discussed in \citet{Bernhard2014}.
The analysis of Fig.~\ref{rapporti} underlines that galaxies fitted by models that adopt SFH$_{con}$, e.g. showing a smoother increase of the SFR compared to the SFR$_{ris}$ and SFH$_{dec}$, represent the majority of the objects at any cosmic time.
At early times their percentage decreases and  the differences between the  three families become less pronounced.
This suggests  that it is  important to allow different types of SFHs, particularly at  higher redshift.

13\% of our sample is made of  galaxies fit by models with declining
SFH and thus high star forming efficiency and low infall time scale (SFH$_{dec}$, green points). Such objects have a history of rapidly mass 
build-up followed by a fast declining phase of star formation, and could  
likely be  the progenitors of quiescent, massive galaxies. These galaxies show, on average, a small amount of dust content and older ages, when compared to galaxies fitted by
SFH$_{ris}$.

\begin{figure}
\begin{sideways}
   \centering
     \includegraphics[width=0.40\textwidth]{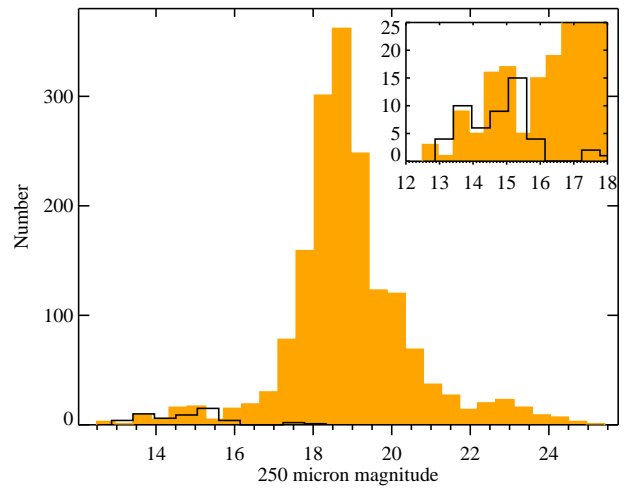} 
     \end{sideways}
     \caption{The  distribution of the 250 $\rm \mu$m magnitudes in the COSMOS field.
      The yellow histogram refers to the magnitudes  estimated by the SED fitting procedure, while the solid black line shows the distribution of the magnitudes observed by Herschel.}   
       \label{isto_250}
   \end{figure}

Another interesting result is the presence of some 
galaxies showing high    
SFR and dust attenuation, which was not evident in \citet{Tasca2015}, also based on the VUDS sample.
This could be due to the fact that in this
paper we allowed extinctions up to $\sim$ A$_{V}$=4, while in \citet{Tasca2015}
the highest extinction allowed corresponds to an A$_{V}$ less than 2 (E(B-V)=0.5). 
It is not surprising that we find these highly star forming galaxies: objects for which the SFR is estimated to be >100 M$_{\odot}$/yr have, for instance,  been detected in the sub-mm or mm bands \citep[e.g.,][]{Toft2014}.
At z$_{spec}$>3 and SFRs>100 M$_\odot$/yr we find a surface density of 0.02/arcmin$^{2}$, that 
can be compared with the expected density of dusty UV 
selected galaxies of 0.23/arcmin$^{2}$ \citep{Mancuso2016}. 
However, given the complex selection function of the VUDS galaxies, we can only state that
these galaxies could be reasonable candidates for observations in the sub-mm/mm
bands with the caveat that the number of galaxies with high SFR increases
as higher attenuations are allowed in the SED fitting procedure.

The COSMOS field has been observed by Herschel and we can compare the 250$\rm \mu$m magnitudes estimated by the best fit models with the observed magnitudes of the galaxies detected by Herschel.
Fig.~\ref{isto_250}  shows the distribution of the 250$\rm \mu$m magnitudes estimated by the SED fitting procedure. The distribution is
peaked at around 19 mag with a bright and a lower fainter tail. The bright tail corresponds to galaxies with the higher SFR and show magnitudes
of the same order of the ones detected by Herschel, shown with a black solid line in Fig.~\ref{isto_250}. The inner histogram shows the zoom of the brightest region.
The flux comparison shown in the histogram is statistical.  
The correspondence is not perfectly 1:1, because for many galaxies the SED fitting procedure shows a magnitudes fainter than the observed one. This could hint to the fact that the Herschel 250$\rm \mu$m fluxes in many cases are due to multiple components \citep{Scudder2016}.

\section{Summary}

In this work we take advantage of VIMOS Ultra Deep Survey (VUDS), the largest spectroscopic survey in the
redshift range 2 $\leq$ z$_{\rm spec}$ <6.54 \citep{LeFevre2015} and we explore the star formation properties of  2995 galaxies with 
reliable spectroscopic redshifts z$_{\rm spec}$ $\geq 2$.
The SED fitting was performed with the GOSSIP+ tool \citep{Thomas2016},
 which allows the simultaneous fitting of 
spectra and photometric data. 
As theoretical models, we built a catalog of galaxies from  SSPs presented
in \citet{Cassara2013} weighted with a set of 28 different SFHs,
capable of reproducing the properties of disks and bulges in the local Universe.\\
We focus on the relation between M$_{*}$ and SFR, and we adopt different assumptions about star formation histories in order 
to investigate their effects on the SFR-M$_{*}$ plane and on the scatter of
this relation. 
We found that the main cause of the scatter in the relation between stellar masses
and SFRs is the contemporary adoption of different SFHs when performing the SED fitting.
These results are in substantial agreement
with \citet{Schaerer2013} for Lyman Break Galaxies at z>3 and \citet{DeBarros2014}, who showed that there is a correlation between the location
where a galaxy ends up in the SFR-M$_{*}$ plane and the type of SFH used in the
best-fitting procedure.
We obtain the same results using a variety of SFHs, which have been left as free parameter for the SED fitting procedure.\\
The effect of the assumed forms of SFHs have been also studied by \citet{Lee2010}, and their analysis has revealed 
that the assumption about SFHs can significantly bias the inference about stellar population parameters,
in particular ages and SFRs.
The recovery of the  properties of galaxies by means of  SED fitting process is indeed strongly influenced
by the choice of the SFHs, and this can cause strong biases in the determination \textit{a posteriori} of their physical properties.
Currently the data are insufficient to distinguish between simple SFHs 
(e.g.  SFHs exponentially declining, rising and constant, SFHs which are unlikely to capture the
full diversity and complexity in the SFHs in galaxies \citep{Reddy2012} and more complicated ones.
The basic conclusion from our analysis is the importance of using a wide range of SFHs when fitting
the SEDs of galaxies at high redshift to derive stellar masses and SFHs, in order to  interpret 
both the relation between these physical quantities and its intrinsic scatter.

 \begin{acknowledgements}

The authors would like to thank the anonymous referee whose criticism has helped to improve the quality of the manuscript.
DM gratefully acknowledges LAM hospitality during the initial phases of the project.
LPC acknowledges the developers of the software TOPCAT (http://www.starlink.ac.uk/topcat/).
LPC thanks Paolo Franzetti for his help and invaluable discussions.
LPC acknowledges EU FP7 funded project DustPedia.
DustPedia is a collaborative focused research project supported by the European Union under the Seventh Framework Programme (2007-2013) call (proposal no. 606847), with participating institutions: Cardiff University, UK; National Observatory of Athens, Greece; Ghent University, Belgium; UniversiteParis Sud, France; National Institute for Astrophysics, Italy and CEA (Paris), France.
This work is supported by funding from the European Research Council Advanced Grant ERC-2010-AdG-268107-EARLY and by INAF Grants PRIN 2010, PRIN 2012 and PICS 2013. 
AC, OC, MT and VS acknowledge the grant MIUR PRIN 2010--2011. 
This work is based on data products made available at the CESAM data center, Laboratoire d'Astrophysique de Marseille. 
This work partly uses observations obtained with MegaPrime/MegaCam, a joint project of CFHT and CEA/DAPNIA, at the Canada-France-Hawaii Telescope (CFHT)
 which is operated by the National Research Council (NRC) of Canada, the Institut National des Sciences de l'Univers of the Centre 
National de la Recherche Scientifique (CNRS) of France, and the University of Hawaii. This work is based in part on data products produced at 
TERAPIX and the Canadian Astronomy Data Centre as part of the Canada-France-Hawaii Telescope Legacy Survey, a collaborative project of NRC and CNRS.

\end{acknowledgements}
\bibliographystyle{aa} 
\bibliography{Cassara2016_astroph} 
\begin{appendix}
\section{Models: single stellar populations}\label{app:SSPs}
A detailed description of both features and novelties of the SSPs models
can be found in \citet{Cassara2013}, here we give a short summary of their
main characteristics:

\begin{itemize}
\item The \citet{Bertelli1994} library up to the end of the E-AGB and a state-of-art treatment for the AGB phase 
for the intermediate and low mass stars \citep{Weiss2009} have been adopted;
\item The stellar models of the \citet{Bertelli1994} library are
those of \citet{Alongi1993,Bressan1993,Fagotto1994a,Fagotto1994b,Fagotto1994c,Girardi1996} and were calculated with the
Padua stellar evolution code;
\item The models of \citet{Weiss2009} were
calculated with the Garching Stellar Evolution Code \citep{Weiss2008};
\item All evolutionary phases, from the
zero-age main sequence to the start of the TP-AGB stage or central
C ignition are included;
\item The age ranges from 0.005 to 20 Gyr (65 values of ages in total);
\item The range of wavelength extends from 0.1 to 1000 $\mu$m (rest-frame);
\item Five values of metallicity (Z=0.0004, 0.004, 0.008, 0.02 - the solar value - and 0.05) have been employed;
\item The primordial He-content is $Y=0.23$ and the enrichment law is $\Delta Y / \Delta Z=2.5$; given the metallicity $Z$ 
and the helium content $Y$, the \citet{Grevesse1993} tabulations of the abundances of heavy 
elements that compose the total metallicity have been considered.
\end{itemize}

\noindent The local effect of absorption/emission due to the  molecular clouds
mimics the effect of the dust around the stellar population. It is calculated applying 
the ray tracing technique, which considers a dust component emitting in the IR and 
fully conserves the energy balance between the dust-absorbed stellar emission in the UV-optical
range and its re-emission in the IR \citep{Takagi2003,Piovan2006a}.

\subsection{The treatment of dust extinction and re-emission: the ray tracing technique}\label{app:dust}

The best way to include attenuation from interstellar dust in the modelling of stellar
populations in galaxies is to solve the radiative transfer equation, in order to build physical 
and self-consistent galactic SEDs \citep{Panuzzo2007,Buat2012,Conroy2013}.\\
With the availability of mid and far-IR data for large samples of galaxies, new codes that 
combine stellar and dust emission on the basis 
of the balance between the stellar luminosity absorbed by dust
and the corresponding luminosity re-emitted in the IR are emerging \citep[just to mention a few:][]{dacunha2008,Noll2009}.\\
These codes make use of attenuation laws, with the exception of those which include a 
full radiation transfer treatment. \\
The most popular attenuation curve in use is the Calzetti law \citep{Calzetti1994,Calzetti2000},
built for local starburst galaxies, which is  also used to include dust attenuation 
by fixing the shape of the attenuation curve and fitting for the normalization.
It is worth underlining that the Calzetti law does not exhibit
the bump at 2175 $\AA$: on the contrary, in all radiative transfer
dust models the expectation is that normal star forming galaxies
should show evidence for this dust feature provided
that the underlining grain population is similar to that of MW or
LMC, even if the question is still under debate, because of the
paucity of rest frame UV spectra of star forming galaxies \citep{Conroy2013}.\\
\citet{Noll2009} presented stacked rest-frame UV
spectra of z $\sim$ 2 star forming galaxies and found strong evidence
for the presence of the 2175 $\AA$ feature with a strength slightly
weaker than observed in the MW extinction curve, and \citet{Wild2011}, in the UV, find evidence for a bump in the attenuation
curve of spiral galaxies at 2175 $\AA$ \citep[see also][]{Conroy2010a}. \\
This feature could have a significant impact on the interpretation of high redshift galaxies
\citep{Gonzalez2013,Mitchell2013}.\\
To account for the effect of extinction and re-emission by dust, we have adopted the
the ray-tracing radiative transfer code \citep[see][for an exhaustive treatment
of the topic]{Takagi2003,Piovan2006a}.
The radiative transfer code considers the cloud (MC hereafter) as a spherical object 
with dust, gas and stars having the same spatial distribution across the whole region.
The equation of radiative transfer (needed because the high density in the regions of star formation 
leads to a very high optical depth, also for IR photons) is solved  
along a set of rays traced throughout the inhomogeneous spherically symmetric source
and the effect of absorption and scattering of the light due to the dust of the molecular
clouds  is taken into account.\\
The spherical symmetry of the problem gives the possibility of calculating the specific intensity of the 
radiation field at a given distance from the center of the MC by averaging the intensities of all rays passing through that
point \citep{Band1985}.\\
Once specified the SSP illuminating dust, typical parameters are:
\begin{itemize}
\item \textbf{$R$}: scale radius of the cloud: more or less compact MCs;
\item \textbf{b$_{\rm c}$}: the abundance of very small carbonaceous grains (VSGs, e.g. PAHs and very small graphite grains), 
influencing the MIR emission in particular;
\item \textbf{Ion}: the ionization model of PAHs; 
\item \textbf{$\tau_{\rm V}$}: the optical depth of the cloud at a fixed wavelength (for example, in the V-band).
\end{itemize}

The key parameter is the optical depth of the MCs.
Its effect on the spectrum emitted by the cloud is that the higher the optical depth, the greater amount of energy 
is shifted toward longer wavelengths. The amount of IR re-emitted luminosity first quickly increases 
at increasing optical depth, then becomes less sensitive to $\tau_{\rm V}$ and  tends to flatten out for
high $\tau_{\rm V}$ \citep{Piovan2006a}.\\

The database of dusty SSPs has been calculated considering:
\begin{enumerate}
\item \textbf{$Z$}: five values of metallicity and 40 values of age for each metallicity;
\item \textbf{$R$}: scale radius of the cloud normalized to the SSP mass. $R$ links the mass of the sources of radiation to
the dimension of the cloud and its effect is to change the position of the FIR peak due to dust emission.
An ideal MC scaled with larger value of $R$ will have a lower temperature profile of grains because of
the bigger dimensions. 
\item \textbf{$\tau_{\rm V}$}: the optical depth of the cloud in the V-band:
20 values of $\tau_{\rm V}$:  0.01, 0.03, 0.05, 0.08, 0.1, 0.3, 0.5, 0.8, 1, 1.3, 1.5, 1.8, 2, 2.3, 2.5, 2.8, 3, 3.3, 3.5, 4 have been adopted;
\item \textbf{b$_{\rm c}$}: the abundances of VSGs: this parameter is related to the extinction curve
\citep{Piovan2006a} and is kept fixed to its maximum value (except for the SMC extinction curve where
only one value is available);  
\item \textbf{Ion}: together with b$_{\rm c}$, this parameter affects the PAHs emission.
we use the ionization profile calculated in \citet{Li2001} for the diffuse ISM of the MW;
\item \textbf{Extinction curves}: we use three extinction curves (MW, LMC and SMC). This parameter is related 
to the metallicity of the stellar populations whose radiation will be reprocessed from the MC. 
\end{enumerate}

These dusty SSPs are the seeds of theoretical models of galaxies.
The last point to underline is that the adopted choices for $\tau_{\rm V}$, b$_{\rm c}$, the ionization model of PAHs and the extinction curves
have been applied in order to reduce the number of free parameters. As widely discussed in \citet{Piovan2006a}, the ideal case would be to set up a library of SSPs covering an ample range of optical depths, at varying the mass and the ratio of the molecular clouds.
For our purpose, it is adequate to fix the parameter controlling the MIR and the FIR emission \citep{Takagi2003,Piovan2006a}
while varying the values of the optical depths, key parameter of the radiative transfer problem, for which an ample range of values has been allowed.

\section{Composite Stellar Populations (CSPs models)}\label{app:galaxies}

A model galaxy with a certain star formation history SFH, SFR(t) and 
chemical enrichment history, $Z(t)$, can be modeled as the convolution of SSPs of different age, weighted by the SFR 
and chemical composition.
We started from the database of dusty SSPs presented before and we calculated a set of theoretical galaxies 
considering 28 SFHs with different ratios of infall time $\tau_{\rm infall}$ to star formation efficiency $\nu$, 
in order to model various morphological types \citep{Buzzoni2002,Piovan2006b}.
The main features and parameters involved in the galactic models are:

\begin{itemize}
\item\textbf{the galactic mass M(t$_{\rm Gal}$)}: in the infall models, it represents the asymptotic
value which the inflowing material reaches at the final galaxy age. The age
t$_{\rm Gal}$ is one of the input of the spectro-photometric code, and it is fixed once
the cosmological framework, and in particular the age of galaxy formation, is
established. The galactic mass is expressed in 10$^{12}$ $\times$ M$_{\odot}$;
\item \textbf{k}: the exponent $k$ of the Schmidt law (see Appendix~\ref{app:SFHs}): for all models, $k$ = 1;
\item \textbf{$\nu$}: the efficiency $\nu$ of the star formation rate is related to 
the galactic mass when simulating bulge galaxies, in order to reproduce the observed
trend of less massive galaxies that  keep forming stars over a longer period with respect to with respect to 
more massive ones \citep{Tantalo1996,Thomas2005}, while 
$\nu$ = 0.50 and 0.35, following \citet{Portinari1998} and \citet{Cassara2015} in case of disk models;
\item \textbf{$\tau_{\rm infall}$}: the infall time scale. $\tau_{\rm infall}$ = 0.3 Gyr for bulge models and 
$\tau_{\rm infall}$= 0.70, 1, 3, 5 in order to account for a different time-extended
star formation in disk galaxies \citep{Portinari1999,Portinari2000}. 
\item \textbf{slope and $\zeta$ of the IMF}: the slope is kept constant at the
classical value of Salpeter, whereas for $\zeta$ we adopt three values, e.g. $\zeta$=0.50, 0.39 and 0.35. 
As first reported in \citet{Bressan1994} and subsequently discussed in \citet{Cassara2015}, the use of the Salpeter law for the IMF require a proportionality constant.  
This constant is fixed by imposing that the fraction $\zeta$ of the IMF mass comprised between 1 M$_{\odot}$ (the minimum mass whose age is comparable to the age of the Universe) 
and the upper limit, i.e. the mass interval effectively contributing to nucleosynthesis. This parameter affects the
metallicities of the galactic models.
\end{itemize}

The constraint of relating the efficiency of the star formation rate to the galactic mass for the bulge galaxies, when building the chemical evolution of the models, comes from the following considerations.
The chemical model in use is a static one, e.g. the formation of the galaxy is simulated by the collapse of primordial gas in presence of the dark matter in a simple fashion. The model galaxy is conceived as a mass point \citep{Chiosi1980} for which no information about the spatial distribution of stars and gas is available. This kind of classical chemical simulation adopts the star formation as an input. In order to reproduce the observed trend of maximum duration of star formation decreasing with the galactic mass we need to relate the efficiency of the star formation to  M(t$_{\rm Gal}$). For a detailed discussion of this topic we refer to  \citet{Tantalo1996,Piovan2006b,Cassara2012,Cassara2015}.

\begin{table}[!ht]
\centering
\begin{tabular}{|l|l|l|l|}
\hline
M(t$_{\rm Gal}$) & $\nu$ & $\zeta$ & $\tau_{\rm \rm infall}$ \\
\hline
5.00 & 14.00 & 0.50 & 0.30 \\
\hline
3.00 & 12.00 & 0.50 & 0.30\\
\hline
1.00 & 7.20 & 0.50 & 0.30\\
\hline
0.50 & 5.20 & 0.50 & 0.30\\
\hline
0.10 & 3.00 & 0.50 & 0.30\\
\hline
5.00 & 14.00 & 0.39 & 0.30\\
\hline
3.00 & 12.00 & 0.39 & 0.30\\
\hline
1.00 & 7.20 & 0.39 & 0.30\\
\hline
0.50 & 5.20 & 0.39 & 0.30\\
\hline
0.10 & 3.00 & 0.39 & 0.30\\
\hline
1.00 & 7.20 & 0.50 & 0.70\\
\hline
3.00 & 12.00 & 0.50 & 0.70\\
\hline
0.50 & 5.20 & 0.50 & 0.70\\
\hline
1.00 & 0.50 & 0.35 & 1.00 \\
\hline
1.00 & 0.50 & 0.35 & 3.00 \\
\hline
1.00 & 0.50 & 0.35 & 5.00 \\
\hline
1.00 & 0.50 & 0.50 & 1.00 \\
\hline
1.00 & 0.50 & 0.50 & 3.00 \\
\hline
1.00 & 0.50 & 0.50 & 5.00 \\
\hline
1.00 & 0.35 & 0.50 & 1.00 \\
\hline
1.00 & 0.35 & 0.50 & 3.00 \\
\hline
1.00 & 0.35 & 0.50 & 5.00 \\
\hline
1.00 & 0.50 & 0.50 & 0.50 \\
\hline
1.00 & 0.50 & 0.35 & 0.50 \\
\hline
1.00 & 0.35 & 0.50 & 0.50 \\
\hline
1.00 & 1.00 & 0.50 & 0.50 \\
\hline
1.00 & 1.50 & 0.50 & 0.50 \\
\hline
1.00 & 2.00 & 0.50 & 0.50 \\
\hline
\end{tabular}
\caption{Parameters for galaxy models} 
\label{table_galaxies}
\end{table}

\begin{figure}
\centering
\subfigure
{\includegraphics[width=0.50\textwidth]{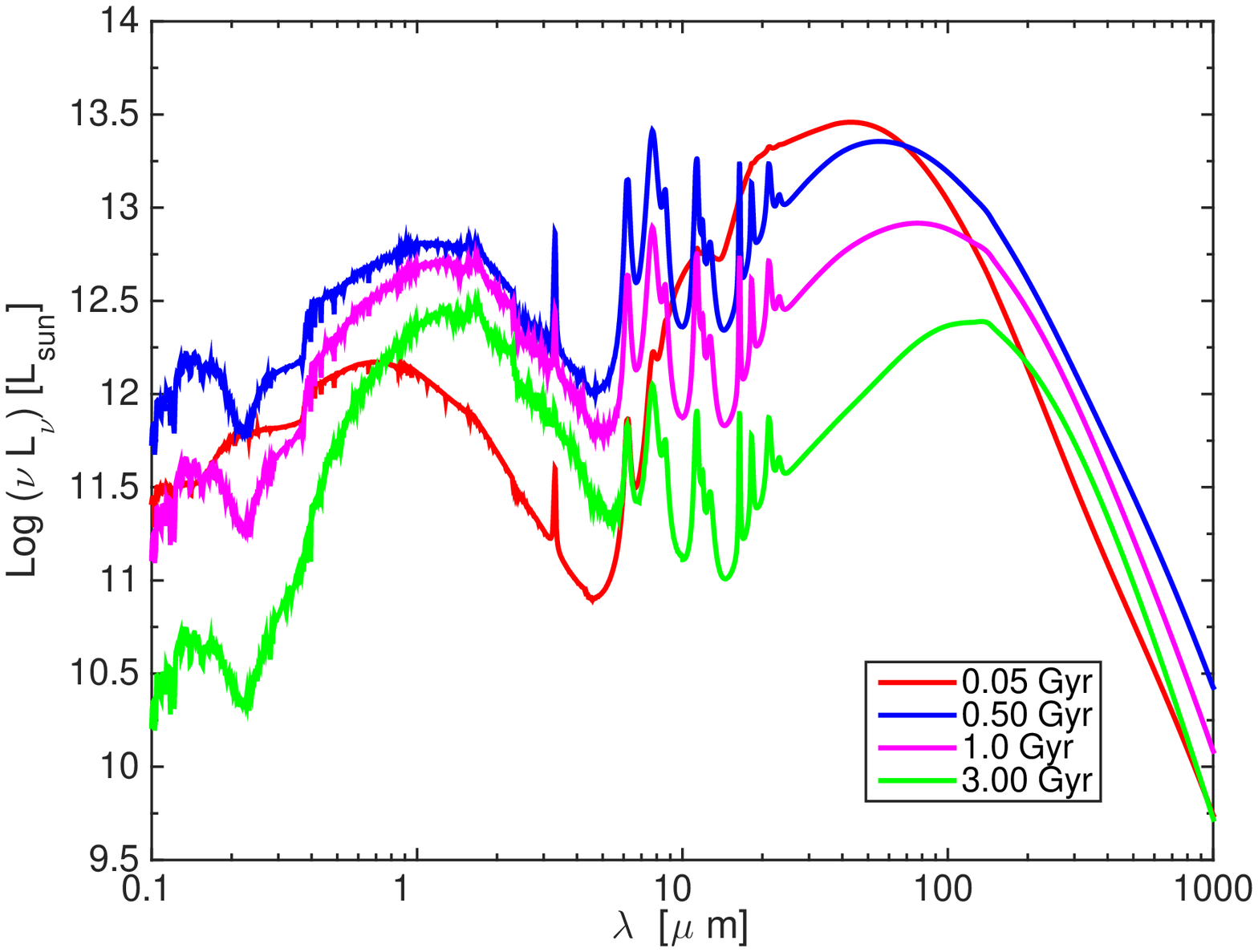}}
\subfigure
{\includegraphics[width=0.50\textwidth]{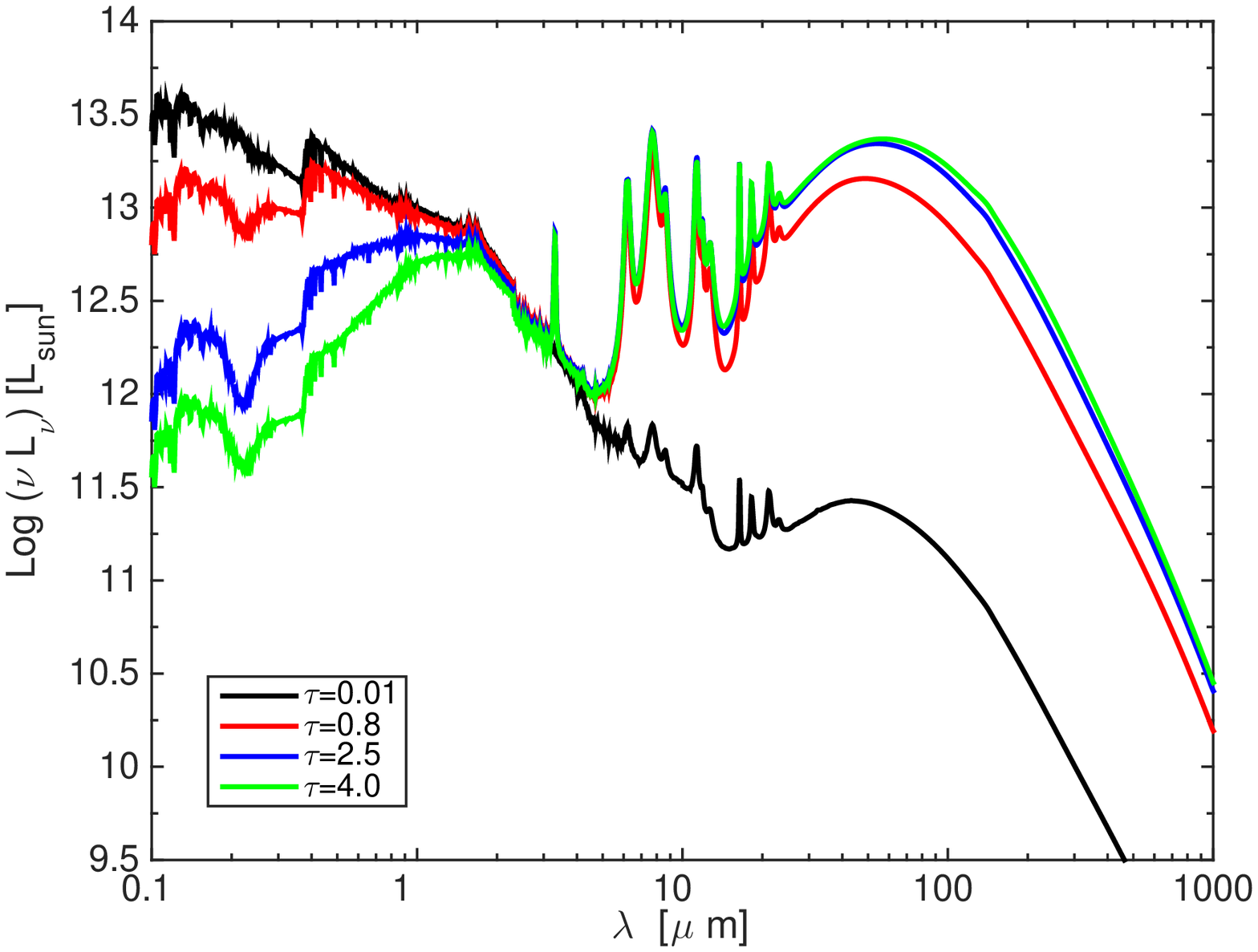}}
\caption{SEDs of theoretical galaxies at different ages and for $\tau_{\rm V}$=2.5 (upper panel)  and with different optical depth of the MC $\tau_{\rm V}$ and or t=0.75 Gyr (lower panel).}
\label{sed_gal}
\end{figure}

\noindent Typical resulting evolutionary galaxy SEDs are shown in Fig.~\ref{sed_gal}: in the upper panel, 
the evolution with the age of the stellar populations, while in the lower panel, SEDs at varying the optical depth of the MCs.\\
The real advantage of a full and physically consistent SEDs which go from the far-UV to the FIR 
is that it reveals important components of a galaxy that are not noticeable in the UV where young, massive 
stars are the dominant flux contributor.\\
Our library of model galaxies consists in $\sim$ 30,000 models: 28 galactic models $\times$ 20 optical depths $\tau_{\rm V}$ $\times$ $\sim$ 55 values of
age for each evolutionary model.\\
It is worth underlining these kind of evolutionary models of galaxies of different morphological 
types are able to reproduce CMDs of local galaxies and the color-redshift evolution \citep{Bressan1994,Piovan2006b,Cassara2015}.

\subsection{Star formation rate/histories}\label{app:SFHs}

The star formation and chemical enrichment histories of this kind of theoretical models of galaxies
are fully described in a number of papers \citep{Chiosi1980,Tantalo1996,Portinari1998,Piovan2006b} and will not be repeated here.
Here we only report the most important points.
The star formation rate i.e. the number of stars of mass $M$ born in the time interval $d$t and mass
interval $d$M, is given by $d$N/$d$t =$\Psi(t)SFR(M)d$M.\\
The rate of star formation SFR(t) is the \citep{Schmidt1959} law adapted to the models, 
SFR(t)= $\nu M_{\rm g}(t)^{k}$, where M$_{\rm g}$ is the mass of the gas at the time $t$.
The parameter $\nu$ and $k$ are extremely important: $k$ yields the dependence of the SFR
on the gas content, while $\nu$ measures the efficiency of the star formation process.
In this type of models, because of the competition between gas infall, gas consumption
by star formation and gas ejection by dying stars, the SFR starts very low, grows to a maximum 
and then declines.\\
The time-scale roughly corresponds to the age at which the star formation
activity reaches the peak value.\\
The last point worth emphasizing is that the galaxy models stand on a robust model of chemical evolution 
that, assuming a suitable prescription for gas infall, IMF, SFR and 
stellar ejecta, provides the total amounts of gas and stars present at any age together with their 
chemical histories, to be used as entries for the population synthesis code
\citep{Chiosi1980,Tantalo1996,Portinari1998,Portinari2000,Piovan2006a,Piovan2006b,Cassara2015}.
\end{appendix}

\end{document}